\documentclass[
    aps,
    prx,
    twocolumn,
    superscriptaddress,
    amsmath,
    amssymb
]{revtex4-2}

\usepackage{graphicx}
\usepackage{dcolumn}
\usepackage{bm}
\usepackage{hyperref}

\usepackage{braket}
\newcommand{\ketbra}[2]{\mathinner{|{#1}\rangle\langle{#2}|}}
\usepackage{tikz}

\usepackage[caption=false]{subfig}
\usepackage{multirow}

\usepackage[ruled,linesnumbered]{algorithm2e}

\usepackage[dvipsnames]{xcolor}

\definecolor{unipdRed}{RGB}{155, 0, 20}
\hypersetup{
    colorlinks,
    urlcolor=unipdRed,
    linkcolor=unipdRed,
    citecolor=unipdRed
}

\usepackage{orcidlink}
\newcommand{\orcidbaronefr}{\orcidlink{0000-0003-2689-8499}}
\newcommand{\orciddaniel}{\orcidlink{0000-0001-7658-3546}}
\newcommand{\orcidilaria}{\orcidlink{0000-0002-3806-2034}}
\newcommand{\orcidsimone}{\orcidlink{0000-0002-8882-2169}}

\newcommand{\affiliateDFA}{\affiliation{Dipartimento di Fisica e Astronomia “G. Galilei”, via Marzolo 8, I-35131, Padova, Italy.}}
\newcommand{\affiliateQCSC}{\affiliation{Padua Quantum Technologies Research Center, Universit{\`a} degli Studi di Padova, Padova, Italy.}}
\newcommand{\affiliateINFN}{\affiliation{INFN, Sezione di Padova, Padova, Italy.}}
\newcommand{\affiliateUlm}{\affiliation{Institute for Complex Quantum Systems, Ulm University, Albert-Einstein-Allee 11, 89069 Ulm, Germany.}}

\begin{document}

\preprint{APS/123-QED}

\title{{Color code thresholds under circuit-level noise beyond the Pauli framework}}%

\author{Francesco Pio Barone\orcidbaronefr}
\email[Corresponding author: ]{francescopio.barone@phd.unipd.it}
\affiliateDFA \affiliateINFN
\author{Daniel Jaschke\orciddaniel} \affiliateUlm \affiliateDFA \affiliateINFN 
\author{Ilaria Siloi\orcidilaria} \affiliateDFA \affiliateINFN \affiliateQCSC
\author{Simone Montangero\orcidsimone} \affiliateDFA \affiliateINFN \affiliateQCSC

\date{\today}

\begin{abstract}
A quantum error correction code is assessed over its ability to correct errors in noisy quantum circuits.
This task requires extensive simulations of faulty quantum circuits, which are often made tractable by considering stochastic Pauli noise models, as they are compatible with efficient classical simulation techniques.
However, such noise models do not fully capture the variety of physical error mechanisms encountered in realistic quantum platforms.
In this work, we extend circuit-level noise modeling beyond the Pauli framework by estimating the threshold of the hexagonal color code under more general noise models. 
Specifically, we consider two representative non-Pauli error channels: a systematic $X$-rotation model that introduces coherent over-rotations, and an amplitude damping channel that captures relaxation processes. 
These models are incorporated at the circuit level into color code circuits using a Tree Tensor Network ansatz.
Our simulations demonstrate that tensor network simulations enable accurate threshold estimation under non-Pauli noise for color codes up to distance $d=7$ (73 qubits). 
Comparing our results with the Pauli twirling approximations of the noise models, we find that coherent over-rotations yield systematically higher error rates, deviating from the Pauli twirling approximation as the code distance increases.
\end{abstract}

\maketitle


Quantum error correction is a foundational component for realizing scalable quantum processors~\cite{gottesman2009introductionquantumerrorcorrection}.
Its primary goal is to robustly encode logical quantum states, making computation reliable despite inevitable errors.
The logical protection is achieved through a Quantum Error-Correcting Code (QECC), which specifies the structure of the logical encoding and determines which errors can be detected and corrected.
While a variety of QECC constructions exist~\cite{gottesman2009introductionquantumerrorcorrection,PhysRevA.86.032324}, any code must be assessed by its ability to reduce logical error rates relative to the underlying physical noise strength. 
A central requirement for any quantum error-correcting scheme is to operate in a regime such that its logical error rate improves when scaling up the QECC.
In many QECC families, the scaling is controlled by the code distance $d$, an integer number proportional to the maximum number of correctable errors.
A QECC can be considered advantageous if increasing $d$ leads to smaller logical error rates. 
However, this improvement is guaranteed only when the physical error strength lies below a critical \emph{threshold}~\cite{doi:10.1126/science.279.5349.342,10.1145/258533.258579}.\\

Although analytical techniques can provide bounds~\cite{Dennis_2002,WANG200331,PhysRevX.2.021004,PhysRevLett.120.180501} and can be refined up to the point of including more complex noise modelizations~\cite{Chubb_2021,Vodola2022fundamental}, numerical simulations remain the primary tool for quantitatively characterizing the performance of quantum error-correcting codes~\cite{PhysRevA.89.022321}.
Indeed, the degree of protection offered by a given code can be evaluated directly by injecting errors into simulated circuits and tracking the resulting logical error rate.
The apparatus of numerical simulations to estimate the threshold relies on the choice of a \emph{noise model}, which describes how the quantum state is altered at the physical level.
The computational complexity of such simulations varies significantly depending on the noise model at hand. 
When noise is restricted to Clifford operations, such as Pauli string errors, efficient simulation is possible using stabilizer-based techniques~\cite{PhysRevA.70.052328,gidney2021stim}.
In contrast, the treatment of non-Clifford noise demands computationally expensive exact simulations.
A common workaround is to employ formal approximations, such as Pauli twirling~\cite{PhysRevA.87.030302}, that map a generic noise model to a distribution of Pauli strings. However, such an approximation may deviate significantly from the true dynamics.
Tensor network methods provide an alternative to exact simulations by approximating the quantum state, with computational cost governed not by the set of gates but by the growth of entanglement in the system~\cite{PhysRevLett.91.147902,jozsa2006simulationquantumcircuits,Seitz_2023}.
\\

In this work, we use Tree Tensor Network~\cite{Seitz_2023} simulations of quantum circuits to investigate the threshold behavior of the hexagonal color code under generic non-Clifford noise conditions. Unlike standard approaches, our method does not rely on formal approximations of the noise models and is accurate up to computations that generate moderate entanglement~\cite{montangero2018introduction}. We focus on two representative non-Pauli error models: systematic single-qubit over-rotations and amplitude damping, the latter capturing open-system relaxation dynamics.

Previous work has demonstrated the feasibility of tensor network methods for threshold estimation beyond Pauli approximations, notably in Ref.~\cite{PaulinTNsurface}. Our approach, however, differs in a key aspect: we apply noise dynamically at the circuit level, inserting error channels throughout the execution of the error correction cycle rather than applying a single layer of noise to a logical state.
This also allows us to simulate multiple rounds of stabilizer measurements and study the QECC in a memory capacity.\\

Our simulations show that tensor network approximations can yield accurate threshold estimates for 2D color codes up to code distance $d=7$ (73 physical qubits), beyond which the complexity of the network - reflected in the growth of the maximum bond dimension - makes the calculations expensive to converge. 
Within this accessible regime, which is already well beyond the scope of exact diagonalization techniques, we find threshold values consistent with expectations for topological codes. 
Moreover, amplitude damping error rates closely match their Pauli-twirled counterparts, while coherent over-rotations produce higher error rates and deviate from the twirling approximation as the code distance increases.

Overall, our study highlights both the practical boundaries and the potential of tensor network methods for simulating quantum error correction protocols in increasingly realistic noise environments. 
Our findings delineate a feasible operational regime ($d=3,5,7$), which may support further investigations into threshold estimation via coherent information~\cite{Colmenarez_2024}, or simulations of small non-Clifford protocols, such as magic state distillation~\cite{rodriguez2024experimentaldemonstrationlogicalmagic}.\\

The paper is structured as follows. 
In Section~\ref{sec:why}, we discuss the motivations that call for simulations beyond the usual Clifford framework.
Section~\ref{sec:problem} briefly formulates the problem of threshold estimation. A broader introduction to the notions of stabilizer quantum error correction is provided for completeness in Appendix~\ref{apx:qec}. 
Section~\ref{sec:methods} reviews the tensor network simulations of quantum circuits with Tree Tensor Networks, and explains a general algorithm for performing open-system simulations via quantum trajectories. 
Section~\ref{sec:results} presents the numerical results, and Section~\ref{sec:conclusion} summarizes our main findings.
In the remaining Appendix~\ref{apx:layout}, we describe heuristics for optimizing the simulation, such as qubit reordering to reduce the maximum bond dimension.

\section{\label{sec:why}Beyond Clifford simulations}

A broad and widely studied class of error-correcting protocols is based on the stabilizer formalism~\cite{gottesman1997stabilizercodesquantumerror}. In this framework, a QECC is defined by a set of stabilizer operators that can be measured to detect and correct errors.
These stabilizers are expressed as Pauli strings, i.e., tensor products of $X$, $Y$, $Z$, and $I$ acting on subsets of qubits.

Pauli strings belong to the Clifford group, a subset of quantum operations that, while not sufficient for universal quantum computation, can be simulated efficiently on classical hardware~\cite{PhysRevA.70.052328}. 
The key to this efficiency lies in tracking the evolution of a group of stabilizer operators that uniquely defines the quantum state, rather than updating the full quantum state vector. 
The stabilizer picture is polynomially efficient as long as the operations remain within the Clifford group.
Applying non-Clifford gates, on the contrary, will increase the dimensionality of the stabilizer basis to track, making this simulation strategy quickly unfeasible in the number of non-Clifford operations, such as T-gates, that are applied~\cite{PhysRevA.70.052328,Bravyi2019simulationofquantum}.\\

In stabilizer quantum error correction, many standard routines - such as initializing logical states $\ket{\bar{0}}, \ket{\bar{1}}$ and performing syndrome measurements - can be implemented entirely with Clifford gates.
As a result, Clifford-based quantum simulators are particularly well-suited for simulating and validating stabilizer codes.
By simulating the encoding, error injection, and recovery processes, one can estimate the resulting logical error rate and thereby quantify the performance of the code under the given noise assumptions.

To retain the computational advantage of efficient simulation, the noise model must also be restricted to operations within the Clifford group.
For instance, a commonly adopted choice is the $n$-qubit depolarizing noise model. 
Whether considering such a restricted noise model is sufficient to validate a QECC against more general classes of errors, such as coherent errors (small unitary rotations of a single qubit) or dissipative errors arising from interactions with an environment, which are commonly encountered in current quantum platforms~\cite{10.3389/fphy.2024.1360080,PhysRevX.13.041022}, remains an open question.\\

On one hand, it has been rigorously shown that testing a code against stochastic Pauli noise is sufficient for assessing its ability to correct arbitrary errors~\cite{PhysRevResearch.4.043052,PhysRevLett.84.2525,PhysRevLett.77.2585}.
Specifically, if a code can correct all $n$-qubit Pauli errors, it can also correct arbitrary errors affecting up to $n$ qubits.
The underlying intuition is that syndrome measurements effectively project general errors onto a probabilistic mixture of Pauli operators, thereby reducing the problem to a discrete set of correctable error patterns.

On the other hand, one may ask whether it is possible to simulate non-Clifford error models while retaining the computational efficiency of stabilizer-based simulations. A common strategy is to approximate arbitrary noise channels with Pauli error models, through techniques known as Pauli twirling and the honest Pauli approximation~\cite{PhysRevA.87.030302,PhysRevA.87.012324,PaulinTNsurface}.\\

At this point, a natural question arises: Is it necessary to go beyond Clifford-only simulations for quantum error correction, particularly in the context of stabilizer codes? 
While Clifford simulations are formally sufficient for validating a code's ability to correct arbitrary errors, there are scenarios in which this paradigm is insufficient.

First, Pauli approximations of noise channels are not always accurate: they can either overestimate the strength of the noise or reduce it to a simplistic depolarizing model, as shown by Ref.~\cite{PaulinTNsurface} for the surface code.
These inaccuracies can affect the estimation of sensitive performance metrics, such as thresholds and resource overheads. 
Second, simulations involving realistic circuit-level noise are valuable for building digital twins of quantum error correction experiments or optimizing codes for specific hardware noise characteristics~\cite{casanova2024findingquantumcodesriemannian}. 
Finally, as quantum error correction progresses toward full-stack implementations, there is increasing interest in simulating universal logical circuits - such as magic state distillation protocols~\cite{PhysRevA.71.022316,rodriguez2024experimentaldemonstrationlogicalmagic} - which necessarily involve non-Clifford gates and lie beyond the reach of stabilizer-only simulations.

\section{\label{sec:problem}Problem formulation}

In stabilizer quantum error correction, the code is specified by a set of $r$ stabilizer operators $\mathcal{S} \equiv \{s_i:i=1,\dots,r\}$, where each $s_i$ is an $n$-qubit Pauli string. The stabilizers are chosen to generate an Abelian group $\mathcal{S}$, meaning that its generators commute with each other: $[s_i,s_j]=0\;\; \forall i,j$. 
The \emph{code space} $\mathcal{Q}$ is defined as the simultaneous $+1$ eigenspace of all stabilizers, that is, $s_i\ket{\psi}=+\ket{\psi}, \; \forall \ket{\psi} \in \mathcal{Q}, \forall s_i \in \mathcal{S}$. Any state $\ket{\psi} \in \mathcal{Q}$ is referred to as a \emph{codeword}.
On the contrary, any state $\ket{\phi}$ for which there exists a stabilizer $s_* \in \mathcal{S}$ such that $s_* \ket{\phi} = -\ket{\phi}$ lies outside the code space and corresponds to a detectable error.

The dimension of the code space is given by $k=n-r$, reflecting the number of logical degrees of freedom. Since the stabilizer group $\mathcal{S}$ does not fully constrain the $n$-qubit Hilbert space, the remaining degrees of freedom are manipulated by \emph{logical operators}, which are Pauli strings that commute with all stabilizers but act nontrivially on the code space. 
Specifically, one can define $2k$ logical operators $\bar{X}_i$, $\bar{Z}_i$ for $i = 1, \dots, k$, satisfying $[\bar{X}_i, s_j] = [\bar{Z}_i, s_j] = 0$ for all $i,j$, and obeying the same algebra as Pauli operators on $k$ qubits: $[\bar{X}_i, \bar{Z}_j] = 0$ for $i \neq j$, and $\{\bar{X}_i, \bar{Z}_i\} = 0$.

The \emph{distance} $d$ of a code is defined as the minimum weight, i.e., the number of non-identity tensor factors, of any nontrivial logical operator. 
The distance quantifies the code's error-handling capabilities: a code of distance $d$ can detect errors acting on up to $d-1$ qubits and can correct errors affecting up to $\lfloor (d-1)/2 \rfloor$ qubits.
A stabilizer code with $n$ physical qubits, $k$ logical qubits, and distance $d$ is denoted as a $[[n, k, d]]$ QECC.\\

\subsection{The hexagonal color code}

The color code is a widely studied candidate for quantum error correction and has been the focus of several experimental implementations across diverse quantum hardware platforms~\cite{Bluvstein_2023,rodriguez2024experimentaldemonstrationlogicalmagic,doi:10.1126/science.1253742,lacroix2024scalinglogiccolorcode}.
From a theoretical standpoint, the color code possesses several properties that make it an advantageous choice against other QECCs~\cite{Kubica_2015}, such as the surface code~\cite{bravyi1998quantumcodeslatticeboundary}.
Notably, the logical Clifford gates $X$, $Z$, $H$, $CNOT$, and $S$ can be implemented transversally~\cite{PhysRevLett.97.180501}, ensuring that errors from one physical qubit propagate to one other qubit at most. 
Moreover, the color code supports the distillation of high-fidelity magic states 
from noisy ones using only Clifford operations~\cite{landahl2011faulttolerantquantumcomputingcolor,Itogawa_2025}.
\\

\begin{figure*}
    \centering
    \includegraphics[width=\linewidth]{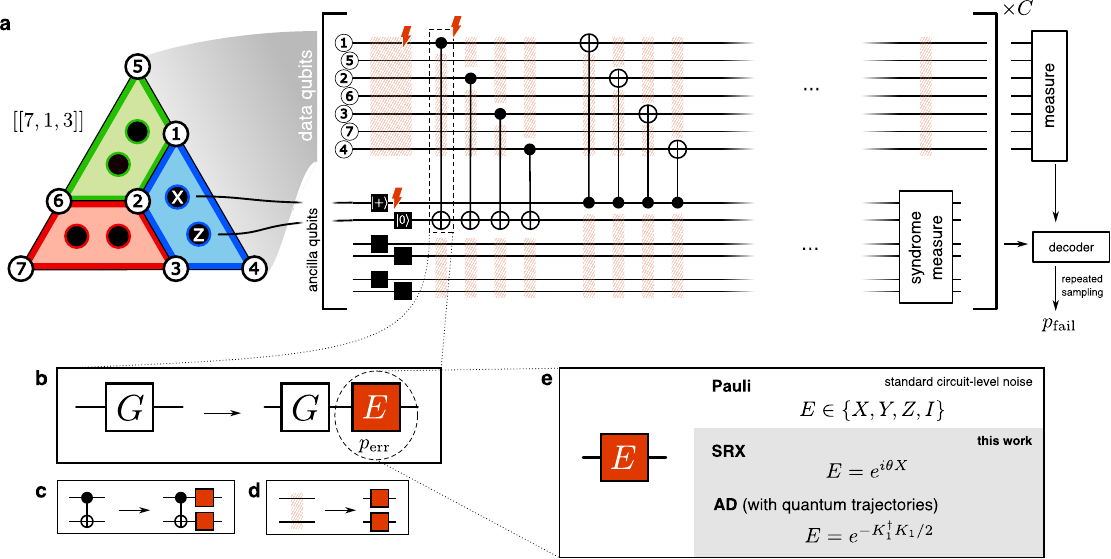}
    \caption{\textbf{Summary of the simulation setup and assumptions of the circuit-level noise model.} 
    \textbf{(a)} The simulated circuit consists of $C$ quantum error correction cycles of the color code. As a reference, we consider the $[[7,1,3]]$ color code, which requires $6$ ancilla qubits for syndrome measurement, leading to a circuit of $7+6$ qubits. The schematic illustrates an example of $s_p^X$ and $s_p^Z$ stabilizer measurements for the blue plaquette; the other plaquettes' syndromes are treated analogously. The scheduling of syndrome measurement operations across all plaquettes follows Ref.~\cite{Lee2025colorcodedecoder}. The striped regions denote idle qubits. 
    \textbf{(b)} Circuit-level noise assumption: every gate is made noisy by executing the ideal gate $G$, followed by some error operation $E$. Moreover, we assume that the error acts on the idle qubits and on the qubit initialization/reset. 
    \textbf{(c)} Gates $G$ acting on more than one qubit are made noisy by applying errors on each of the qubits $G$ acts on. \textbf{(d)} Noise on idle qubits is treated as for a multi-qubit gate. 
    \textbf{(e)} The error operations considered in this work are determined by the SRX (systematic rotations around $X$) and AD (amplitude damping) error models, defined respectively by Eqs.~\eqref{eq:SRX} and~\eqref{eq:AD}.}
    \label{fig:circuitlevel}
\end{figure*}

The smallest code of the hexagonal color code family, $[[7,1,3]]$, is shown on the left of Figure~\ref{fig:circuitlevel}a. The data qubits are labelled from 1 to 7. The stabilizer operators $s_p^X$/$s_p^Z$, defined as products of $X$/$Z$ Paulis, act on the data qubits associated with each of the 3 colored plaquettes $p$.
In the non-fault-tolerant scheme considered here, the stabilizers are measured using sequences of CNOT gates, with the outcomes stored in two additional ancilla qubits per plaquette (represented in the figure by black circles within each plaquette).
This leads to a total number of physical qubits (data plus ancilla) $N_{(d=3)}=7+2\cdot3=13$.
In general, the family of hexagonal color codes is written in function of the distance as $[[(3d^2+1)/4,1,d]]$~\cite{landahl2011faulttolerantquantumcomputingcolor}. 
In this work, we consider the instances $[[19,1,5]]$ and $[[37,1,7]]$, corresponding to $d=5$ and $d=7$, respectively. These lead to total qubit counts of $N_{(d=5)} = 19 + 2 \cdot 9 = 37$ and $N_{(d=7)} = 37 + 2 \cdot 18 = 73$.\\

The measurement outcomes of the stabilizers are referred to as the syndrome. Syndrome measurements provide a non-destructive probe of the code space: any non-trivial syndrome indicates that the state has left the code space and that a recovery operation is required.
The next step in a standard stabilizer QECC pipeline is, therefore, to associate the observed syndrome pattern with a likely error and apply a corresponding recovery operation. This task is performed by a classical \emph{decoding algorithm} (see Appendix~\ref{apx:qec-decoding} for further details).\\

\subsection{Threshold estimation}

Quantum error correction protects the information of $k$ logical qubits by encoding them in $n>k$ physical qubits, but if each physical qubit is equally prone to noise, the total probability of failure may even increase.
Additionally, each syndrome measurement cycle increases the likelihood of error propagation from one qubit to another.
Given the substantial resource overhead, it is of utmost importance to identify the noise regimes in which the code provides a practical advantage.

To this extent, it is common practice to employ simulations that inject errors in the syndrome measurement circuit (controlled by some physical error rate $p_{\mathrm{err}}$ parameter) and use the results from the decoder to estimate the logical error rate $p_{\mathrm{fail}}$~\cite{PhysRevA.89.022321,landahl2011faulttolerantquantumcomputingcolor}.
For QECCs with a non-vanishing threshold, there exists a regime $p_{\mathrm{err}}<\tau$ where the logical error rate is exponentially suppressed in the code distance~\cite{knill1996thresholdaccuracyquantumcomputation}. The task of threshold estimation consists of identifying the threshold $\tau$.
We defer to Appendix~\ref{apx:qec-threshold} for further details.

\section{\label{sec:methods}Methods}

In this section, we describe the setup used to estimate the threshold of the color code, along with the numerical methods employed. 
Our setup largely follows conventions established in previous works in the field~\cite{PhysRevA.89.022321,PhysRevA.86.032324,gidney2021stim}, while extending the protocol to include circuit-level non-Pauli noise simulated using tensor network methods.\\

We simulate quantum circuits composed of $C$ cycles of syndrome extraction for the family of $[[n(d), k=1, d]]$ color codes. 
This allows us to assess the QECC in \emph{memory capacity}, looking at the logical error rate $p_\mathrm{fail}(C)$ after the logical state has gone through $C$ quantum error correction cycles.
The circuit is initialized with all data qubits in the $\ket{0}$ state, which is the +1 eigenvalue $\ket{\bar0}$ of the $\bar{Z}$ operator.
$Z$- and $X$-type stabilizers are measured using two ancilla qubits per plaquette: one initialized in $\ket{0}$ and the other in $\ket{+}$, respectively. When $C\ge2$, the ancilla qubits are measured mid-circuit and subsequently reset for the next error correction cycle. 
The specific CNOT schedule of the syndrome measurement operations follows the optimized schedule from Ref.~\cite{Lee2025colorcodedecoder}, from which we also adopt the decoding algorithm.
The data qubits are measured only at the end of the circuit to determine $\bar{Z}$.\\

The syndrome extraction circuits, sketched in Figure~\ref{fig:circuitlevel}a, implement a non-fault-tolerant form of stabilizer measurement: such circuits do not contain the propagation of errors between ancilla and data qubits, and therefore, single faults do not remain localized, but rather propagate to other qubits.
Moreover, errors accumulate over time in ways that mimic logical faults. 
To reliably detect and correct arbitrary error patterns of weight up to $\lfloor (d-1)/2 \rfloor$, the decoder must be provided with a history of syndrome data spanning a similar temporal depth~\cite{Dennis_2002}. Therefore, we must repeat the syndrome extraction circuit $C \sim \mathcal{O}(d)$ times to accurately estimate the threshold.

Nevertheless, it is also of interest to examine the QECC in a single-shot scenario $C=1$. 
These allow us to isolate the effect of a single round of syndrome measurement, which may still offer practical insights for experimental implementations focusing on the building block of a single round.

\subsection{\label{ssec:circlevel}Modeling non-Pauli circuit-level noise}

An essential aspect of threshold estimation involves injecting noise into the quantum circuit. 
Although there is some flexibility in how noise can be modeled in general quantum circuit simulations, the standard approach in quantum error correction is to include noise by applying the ideal unitary operation followed by a stochastic error gate sampled from a predefined noise model~\cite{Dennis_2002,PhysRevA.86.032324}. 
This approach, known as \emph{circuit-level noise}, captures errors at the level of individual physical operations in the quantum circuit. 
Through the following section, we assume that the noise strength can be controlled by a parameter $p_{\mathrm{err}}$, whose specific interpretation depends on the noise model at hand.
\\

To begin with, a single-qubit gate is made noisy by appending either an $X$, $Y$, or $Z$ gate, sampled with probability $p_{\mathrm{err}}/3$ each, immediately after the ideal gate operation~\cite{Dennis_2002}. 
With probability $1-p_{\mathrm{err}}$, instead, the gate acts ideally.
This noise model is referred to as the \emph{depolarizing noise model}, which corresponds to the depolarizing quantum channel, that is 
\begin{equation*}
    \mathcal{E}_{\mathrm{1QD}}[\rho] = (1-p_{\mathrm{err}})\rho+\frac{p_{\mathrm{err}}}{3}\left(X\rho X+Y\rho Y+Z\rho Z\right)\;.
\end{equation*}
This praxis is generalized to $n$-qubit gates by sampling the error uniformly from nontrivial $n$-qubit Pauli strings.
For instance, a two-qubit gate would be followed by two-qubit errors sampled from the set of 15 elements $\{XI, IX, YI, \dots, IZ, XX, XY, \dots, YZ, ZZ\}$, with probability $p_{\mathrm{err}}/15$ each.
\\

Other circuit operations, such as measurement and initialization operations, are made noisy in a similar stochastic approach. 
For measurements, noise is typically modeled as a readout error: the ideal projective measurement is followed with probability $p_{\mathrm{err}}$ by a bit flip of the returned value. 
For initialization, the qubit prepared in the $\ket{0}$ ($\ket{+}$) state is mistakenly prepared to $\ket{1}$ ($\ket{-}$) with probability $p_{\mathrm{err}}$. Mid-circuit qubit resets are handled similarly to initialization.
In addition, idle qubits can be made noisy by filling the idle locations of the circuit with error operations sampled from $\mathcal{E}_{\mathrm{1QD}}$.

The depolarizing noise applied after an ideal gate, together with the stochastic noisy measurements, faulty initialization, and idle noise, constitutes a standard error model for quantum error correction benchmarks.
Indeed, this model is efficiently simulable using Clifford simulators when applied to Clifford circuits~\cite{Dennis_2002,PhysRevA.70.052328}.
\\

Here, we extend the standard circuit-level noise model to encompass more general, non-Pauli noise processes, such as coherent unitary errors and open-system noise channels.
The former can be implemented straightforwardly by applying a generic unitary gate $E(p_{\mathrm{err}})$, representing the error, right after the ideal gate. 
The incorporation of open-system noise channels, however, requires more careful treatment.
These channels model irreversible dynamics, e.g., dissipation, and generally lead to mixed quantum states.

Finally, although measurement noise could be modeled simply by flipping classical outcomes with some probability, we neglect it in this work, since it does not represent a genuine generalization of the Pauli bit-flip noise. The assumptions of noise modeling are summarized in Figure~\ref{fig:circuitlevel}b-e.
\\

We now introduce two representative noise models that go beyond the standard Pauli picture. The first is the Systematic Rotation around $X$ (SRX) model, a coherent noise channel that applies a fixed single-qubit unitary rotation:
\begin{equation}\label{eq:SRX}
    \mathcal{E}_{\mathrm{SRX}}[\rho] = e^{-i\theta X}\rho e^{i\theta X}\;\;.
\end{equation}
Here, the strength of the noise is determined by the parameter $\theta\in[0,\pi)$, with $\theta = 0$ corresponding to the noiseless case, in which the channel action reduces to the identity. 
This model captures systematic calibration errors, such as over- or under-rotations due to pulse inaccuracies, commonly arising in superconducting or trapped-ion platforms~\cite{10.3389/fphy.2024.1360080,martínezgarcía2021analyticalexperimentalstudycenter,jeanette2025blindcalibrationquantumcomputer}.
Since $\mathcal{E}_{\mathrm{SRX}}$ is a coherent error channel, it can induce coherent logical errors~\cite{PhysRevA.66.032304,Greenbaum_2017}.
Such effects are not captured by stochastic Pauli approximations, and can lead to significant misestimation of logical error rates and thresholds~\cite{PaulinTNsurface,PhysRevLett.121.190501,PhysRevX.11.041039}.
\\

The second error model that we consider is an Amplitude Damping (AD) channel:
\begin{equation}\label{eq:AD}
    \mathcal{E}_{\mathrm{AD}}[\rho] = \sum_{i=0}^1K_i\rho K_i^\dag\quad,
\end{equation}
with Kraus operators
\begin{equation}\label{eq:admodel}
    K_0 = \begin{bmatrix}
    1 & 0 \\
    0 & \sqrt{1-\gamma}
    \end{bmatrix}, \qquad
    K_1 = \begin{bmatrix}
    0 & \sqrt{\gamma} \\
    0 & 0
    \end{bmatrix}.
\end{equation}
Here, $\gamma\in[0,1]$ represents the probability of decaying from $\ket1$ to $\ket0$ and serves as the noise strength parameter.
Amplitude damping is particularly relevant for superconducting platforms~\cite{PhysRevX.13.041022}.\\

More realistic noise models may include correlated coherent errors, such as over-rotations affecting both qubits of a two-qubit gate. In this work, however, we adopt a minimal model in which each fault location consists of a single qubit error injection. This ensures a consistent definition of fault locations across both noise models.

\subsection{\label{ssec:ttn}Emulation of quantum circuits with Tree Tensor Networks}

Tensor networks offer a compact and scalable representation of quantum states by factorizing the statevector into a network of low-rank tensors~\cite{montangero2018introduction,PhysRevLett.91.147902}.
This structured decomposition significantly reduces the memory footprint from exponential in the number of qubits to polynomial in the tensor rank, enabling the classical simulation of quantum systems that would otherwise be intractable.
A key parameter that governs the efficiency and fidelity of such simulations is the \emph{bond dimension} $\chi$, which specifies the maximum allowed size of the shared indices (or bonds) connecting adjacent tensors in the network. 
Notably, the bond dimension effectively constrains the amount of entanglement that can be represented between subsystems of the quantum state, since it sets an upper bound on the Schmidt rank across any bipartition of the system.
Consequently, the accuracy of tensor network simulations is closely tied to the entanglement structure of the state: they are particularly well-suited for simulating states with limited, area-law entanglement. Still, they may become inefficient or inaccurate for highly entangled states.
\\

To maximize the representational power of a tensor network, one must carefully assign the network's physical sites to the qubits of the circuit.
The use of a Matrix Product State (MPS) imposes a linear ordering of the tensors, and therefore of the circuit qubits. Instead, a Tree Tensor Network (TTN) allows for a hierarchical tree structure of the tensors, which can more naturally capture the entanglement geometry of some quantum circuits, especially those that show a clusterable layout~\cite{Seitz_2023,PhysRevB.82.205105}.
The computational cost of applying a two-qubit gate scales with the distance between the two leaf nodes in the tree. For a TTN with $N$ leaves, this distance scales as $\mathcal{O}(\log N)$, yielding a significant improvement over the $\mathcal{O}(N)$ scaling characteristic of MPS-based methods~\cite{Seitz_2023}.
\\

In the context of the present work, the quantum circuits to be simulated are dominated by two-qubit gates (CNOTs) acting on qubits arranged on a two-dimensional lattice. 
Representing such interactions efficiently using a linear tensor network structure, such as an MPS, is challenging due to the 2D spatial locality imposed by the stabilizer operators. 
To address this, we adopt a TTN ansatz, optimizing its structure through a variant of the algorithm proposed in Ref.~\cite{Seitz_2023} (see Appendix~\ref{apx:layout}). The algorithm aims to construct a hierarchical tree structure that clusters qubits according to the connectivity patterns defined by the syndrome extraction circuits, thereby reducing entanglement overhead across the network and improving simulation efficiency.

Our implementation relies on the Quantum Matcha Tea~\cite{qmatchatea,qtealeaves} library, which supports TTNs constrained to a binary tree structure. 
Accordingly, the optimization algorithm shown in Appendix~\ref{apx:layout} is enforcing this constraint.
This optimization yields substantial resource savings - for example, in some cases the required bond dimension is reduced from 2048 to 256 (see Table~\ref{tab:bdexact}).

\subsection{\label{ssec:trajectories}Quantum trajectories}

A commonly employed approach to simulate open quantum systems is the quantum trajectories method~\cite{Jaschke_2018b,bonnes2014superoperatorsvstrajectoriesmatrix,Daley04032014,Jaschke_2018,PhysRevLett.116.237201}. 
Rather than evolving the system as a density matrix, whose size scales quadratically with the Hilbert space dimension, this technique simulates an ensemble of pure-state evolutions. 
These individual trajectories evolve under a non-hermitian effective Hamiltonian, punctuated by quantum jumps that are triggered by a probabilistic criterion.
Averaging over many such trajectories reconstructs the behavior of the open system.
In this work, instead of utilizing an exact representation of the quantum state, we apply the method to a TTN representation of the state.
\\

The Lindblad master equation is the effective equation of motion of a quantum system interacting with a Markovian environment. It takes the form
\begin{equation}
    \partial_t\rho = 
    -i\left[H,\rho\right]
    + \sum_i \bigg( 
    L_i\rho L_i^\dag - \frac12\{L_i^\dag L_i,\rho\} \bigg)\quad,
\end{equation}
where $\rho$ is the density matrix representing the system state, and $H$ is the system Hamiltonian. The set of operators $\{L_i\}$, known as Lindblad or jump operators, encapsulates the effects of coupling to the environment.
The key idea of quantum trajectories is to rewrite the master equation as a stochastic average over pure-state trajectories, $\rho \simeq N_{\mathrm{samples}}^{-1} \sum_i \ketbra{\Psi_i}{\Psi_i}$. Each trajectory is numerically evolved in time according to the effective Hamiltonian
\begin{equation}\label{eq:Heff}
    H_{\mathrm{eff}} = H-\frac{i}{2}\sum_iL_i^\dag L_i\quad.
\end{equation}
The average over many stochastic trajectories yields a result that is equivalent to the solution of the master equation~\cite{Daley04032014}.\\

A critical point to address is how to adapt the formalism of quantum trajectories to gate-based quantum circuit simulations, where the dynamics are governed by a sequence of discrete unitary gates rather than a continuous-time Hamiltonian evolution. 

To this end, consider a Trotterization of the non-Hermitian effective Hamiltonian $H_{\text{eff}}$ into a number of small time steps $\delta t$, chosen to match the number of gates $D$ in the circuit:
\begin{align}
    e^{-iH_{\mathrm{eff}}t} \simeq& \left(e^{-i\left(\frac{-i}{2}\sum_j L_j^\dag L_j \right) \delta t} e^{-iH \delta t}\right)^D = \nonumber\\
    \simeq& \left(e^{-(\delta t/2)\sum_jL_j^\dag L_j} G_D\right) \dots \left( e^{-(\delta t/2)\sum_jL_j^\dag L_j} G_1 \right). \label{eq:gatetrotter}
\end{align}
In the context of quantum circuits, the unitary evolution operator $e^{-iHt}$ corresponds to the sequence of ideal gates $G_D \dots G_2 G_1$. Therefore, the Trotterization can be factored in the product of $D$ terms, as in Eq.~\eqref{eq:gatetrotter}.

This decomposition suggests a natural interpretation: after the application of a gate $G_i$, one can apply a fictitious (non-unitary) gate 
\begin{equation}
    E_{(\mathrm{open-systems})} = e^{-(\delta t/2)\sum_iL_i^\dag L_i}
\end{equation}
that follows the dynamics of the effective Hamiltonian in Eq.~\eqref{eq:Heff}.
To remain consistent with the circuit-level noise picture, where errors are typically introduced immediately after the ideal gate, we apply this fictitious gate only to the target qubits of $G_i$. 
Alternative choices are possible, such as applying the fictitious operation to all qubits, which may represent idle noise.
\\

While gate-based circuits often abstract away gate durations, assuming instantaneous operations, these timescales become relevant when modeling realistic physical systems. 
In experimental platforms, gates are applied over finite time intervals during which the system interacts with its environment.
The result in Eq.~\eqref{eq:gatetrotter} thus allows us to interpret $\delta t$ as the duration over which each gate exposes the system to environmental noise.

Finally, while the Lindblad master equation describes the open-system dynamics through Lindblad operators $\{L_j\}$, the noise models are typically expressed in terms of Kraus operators, such as the AD noise model of Eq.~\eqref{eq:admodel}. 
In that particular case, the mapping between the two formalisms~\cite{Andersson_2007} is straightforward: since $K_0\sim I$, the only relevant jump operator is $L_1 = K_1/\sqrt{\delta t}$.
\\

\section{\label{sec:results}Results}

In this section, we present the numerical results of our study.
To begin with, we discuss in Subsection~\ref{ssec:tnapprox} how the entanglement truncation in the tensor network simulations affects the estimation of the logical error rate.
Threshold estimates are presented and discussed in Subsection~\ref{ssec:thres}, while comparisons with Pauli-twirling approximations are given in Subsection~\ref{ssec:twirl}.

\newcommand{\unitsSRX}{\theta/\pi}
\newcommand{\unitsAD}{\gamma}
\newcommand{\bdexact}{\chi_{\mathrm{exact}}}
\newcommand{\pfail}{p_{\mathrm{fail}}}

\subsection{\label{ssec:tnapprox}Effect of tensor network approximations on the logical error rate}

We examine how the tensor network representation impacts the encoded logical state of the QECC.
Tensor networks achieve compression of a quantum state by truncating the smallest Schmidt coefficients during tensor manipulations.
This approximation effectively limits the amount of entanglement that the ansatz can represent, with the severity of the truncation being controlled by the chosen bond dimension $\chi$.
Understanding the interplay between the bond dimension and the fidelity of the encoded logical state is therefore crucial, as it directly impacts the accuracy and reliability of subsequent simulations.
\\

We perform numerical simulations in absence of physical noise, varying the bond dimension $\chi$ to systematically assess how the approximation affects the encoded logical state.
We define $\bdexact$ as the bond dimension required for the TTN ansatz to represent the quantum state exactly, i.e., without truncating any Schmidt coefficients throughout the simulation of the circuit.
Simulations with $\chi=\bdexact$ yield exact results, as no information is lost during tensor operations.
The regime of practical relevance lies in the interval $\chi /\bdexact \in (0,1)$, where compression induces approximation errors.

To better isolate the effect of the bond dimension, we deliberately disable all TTN layout optimizations (discussed in Appendix~\ref{apx:layout}).
This choice increases the required bond dimension to saturate the fidelity, $\bdexact$, thereby providing a wider range of accessible sub-optimal values $\chi < \bdexact$ for sampling.
\\

\begin{figure}
    \centering
    \includegraphics[width=\linewidth]{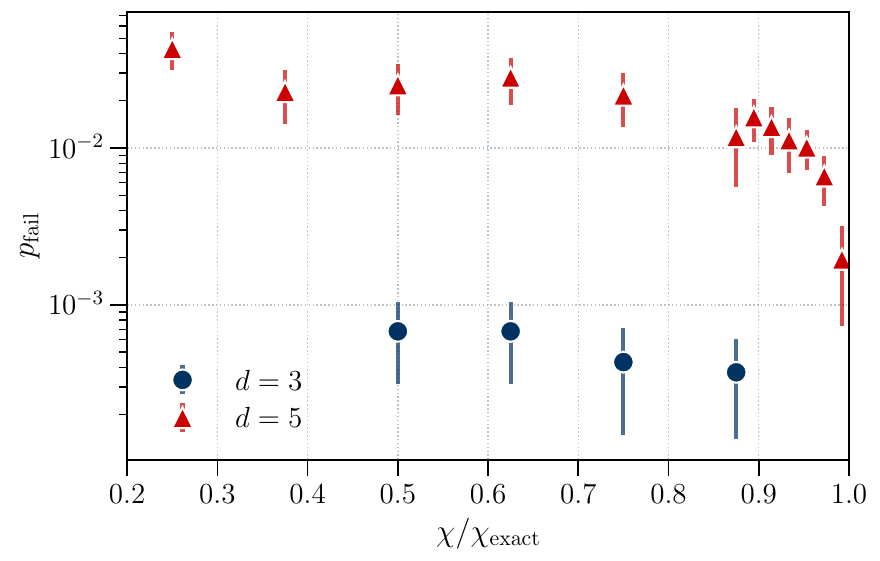}
    \caption{\textbf{Logical error rate induced by the TTN representation.} We report the logical failure rate $\pfail$ as a function of the bond dimension $\chi$, normalized to the exact value $\bdexact$ required for exact TTN simulation in the absence of physical noise.
    For $d=3$, the data is sparse due to the relatively small value $\bdexact^{(d=3)} = 8$, whereas for $d=5$, with $\bdexact^{(d=5)} = 512$, the finer grid in the x-axis allows a more detailed exploration of the sub-exact regime. 
    In the high-cut regime $\chi\ll\bdexact$, the simulations return zero logical error within numerical precision, which is an artifact of extreme truncation of the TTN state.
    The regime of $\chi=\bdexact$ is trivial, as no truncations are executed through the simulation and the logical failure rate vanishes, as expected.}
    \label{fig:tnaccuracy-cut}
\end{figure}

The results of these simulations are presented in Figure~\ref{fig:tnaccuracy-cut}, where we report the logical failure rate $\pfail$ obtained after decoding the syndrome bits.
For a fixed code distance $d$, we observe that $\pfail$ remains nearly constant over a wide range of sub-optimal cuts, $\chi/\bdexact<1$.
This indicates the presence of a plateau region, where moderate truncation affects the logical error rate within the same order of magnitude. Notably, the plateau values differ between $d=3$ and $d=5$. 
This is due to the fact that larger codes involve more qubits and more highly entangled subsystems, increasing the likelihood that TTN truncations discard relevant information. 
The data for $d=5$ reveals that the logical failure rate decreases at convergence $\chi \to \bdexact$.

Overall, a sub-optimal cut $\chi<\bdexact$ introduces a form of ``background'' logical failure rate, which can be estimated by simulating the circuit in the absence of physical noise.\\

In the extreme compression limit $\chi\to1$, the estimated logical failure rate vanishes.
In this regime, we observe a strong bias towards trivial measurement outcomes, consistent with the tensor-network representation being reduced to an effectively product-state manifold.
The product state $\ket{0}^{\otimes n}$ has trivial Z-stabilizer syndromes and is a +1 eigenstate of the $\bar{Z}$ observable. Consequently, in a logical-$\ket{0}$ memory experiment, a decoder presented with such a state would predict the correct value of the logical observable.
In other words, in the high-cut regime ($\chi\ll\bdexact$), tensor truncations can produce false positives by collapsing the output state to a trivially decodable codeword.\\

For intermediate values $\chi\lesssim\bdexact$, the truncation suppresses correlations and effectively removes nontrivial syndrome structure, resulting in a background logical error rate induced by the compression.
We conjecture that this error rate adds to the failure rate induced by physical noise. 
Therefore, simulations in this regime are only meaningful if the logical failure rate due to truncation is much smaller than that caused by the physical error model.
Choosing $\chi = \bdexact$ ensures exact simulation.

\subsection{\label{ssec:thres}Thresholds}

\begin{figure}
\subfloat{%
  \includegraphics[width=\linewidth]{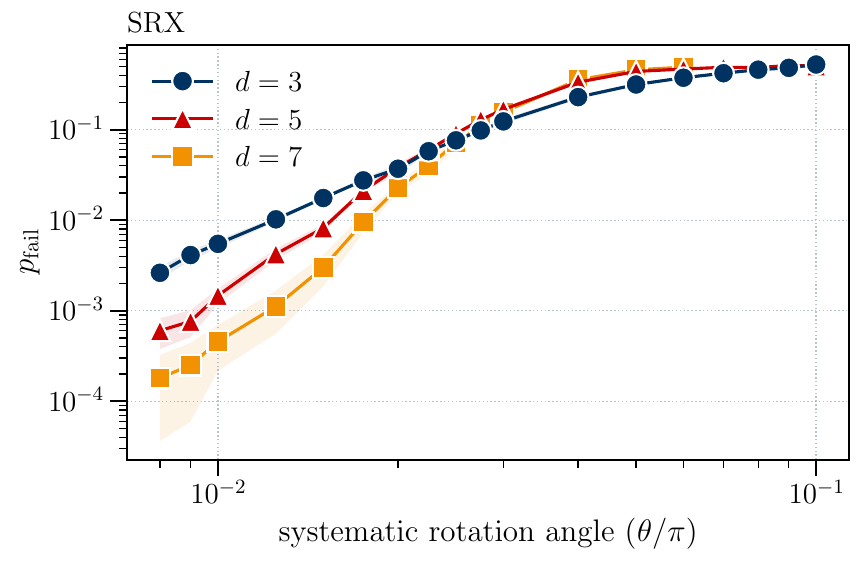}
}\hfill
\subfloat{%
  \includegraphics*[width=\linewidth]{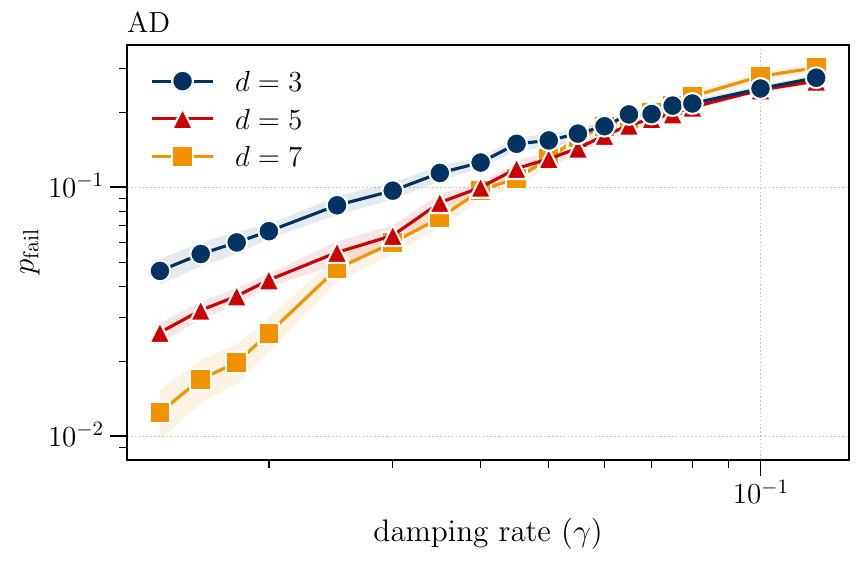}
}
\caption{%
    \textbf{Threshold estimation for one error correction cycle ($C=1$).}
    The top panel corresponds to the SRX noise model, while the bottom panel shows results for the AD model.
    Each data point represents the logical failure rate $\pfail$ estimated from $10^4$ independent decoding trials, where each trial involves a full circuit simulation with injected physical noise.
    Shaded regions around each curve delimit the 99\% confidence intervals, computed assuming binomial statistics for the decoder outcomes (success or failure).
\label{fig:T1-SRX+AD}%
}
\end{figure}

\begin{figure}
\subfloat{%
  \includegraphics[width=\linewidth]{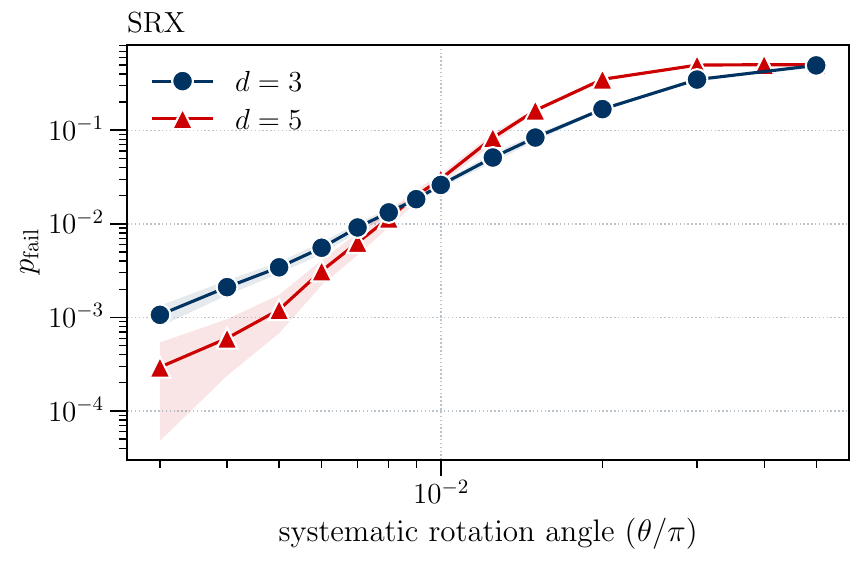}
}\hfill
\subfloat{%
  \includegraphics*[width=\linewidth]{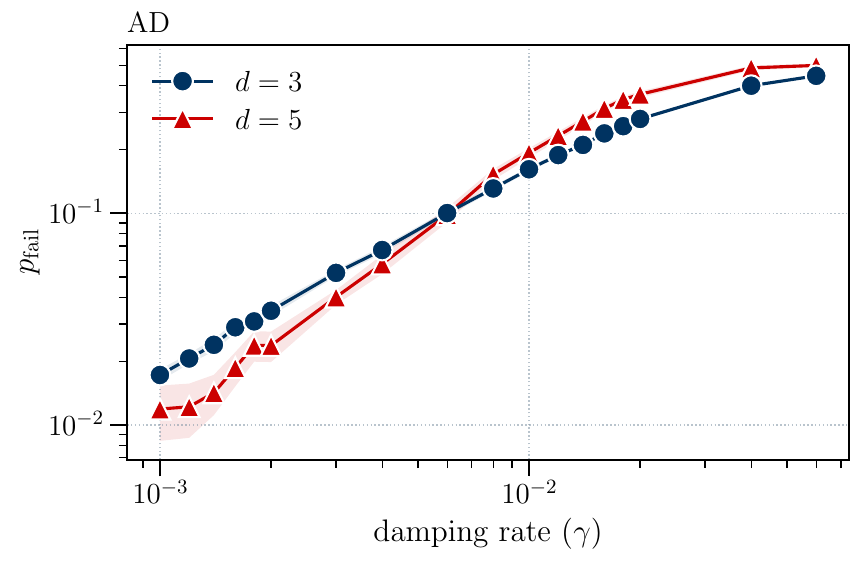}
}
\caption{%
    \textbf{Threshold estimation for repeated error correction cycles ($C=d$).}
    The evaluation of the logical failure rate $\pfail$ follows the same procedure as described in Figure~\ref{fig:T1-SRX+AD}, based on $10^4$ decoding trials per data point with binomial confidence intervals.
\label{fig:Td-SRX+AD}%
}
\end{figure}

Threshold curves are obtained by sampling the logical failure rate $\pfail$ across a range of values of the physical noise strength parameter $p_{\mathrm{err}}$.
For the SRX model, the noise parameter is the rotation angle $\theta$, which we express in units of $\pi$.
For the AD model, the error strength is defined in terms of the damping rate $\unitsAD$.

To estimate the thresholds, we simulate error-correction circuits with physical noise, enabling the TTN qubit layout optimization discussed in Appendix~\ref{apx:layout}. 
Unless otherwise stated, all simulations are performed in a regime of bond dimension $\chi$ that ensures convergence and produces valid results for the task of threshold estimation.
Importantly, these simulations go well beyond the range accessible to brute-force statevector methods. For instance, a color code circuit of distance $d=5$ already requires the simulation of 37 qubits, while for distance $d=7$ the number rises to 73 qubits. The latter size is outside the limits of exact statevector simulations.\\

The results are presented in Figure~\ref{fig:T1-SRX+AD}, which shows the logical failure rate after a single quantum error correction cycle ($C=1$), and in Figure~\ref{fig:Td-SRX+AD}, which reports the behavior after a number of cycles that matches the code distances ($C=d$).
In both cases, the expected threshold behavior is observed, as the points belonging to each set of distances are orderly arranged in the vertical axis of the plots. Below threshold, larger distances suppress $\pfail$; above threshold, the ordering is inverted.

Thresholds can be estimated by looking at the crossing points of logical failure curves for different code distances.
However, more robust values are obtained through a fitting procedure using the ansatz function
\begin{equation}\label{eq:ansatz-tau}
    \pfail = A + Bx + Cx^2 \quad s.t.\quad x =(p_{\mathrm{err}}-\tau)d^{1/\nu}\;,
\end{equation}
where $\tau$ is the threshold and $\nu$ is the critical exponent~\cite{Watson_2014,english2024thresholdspostselectedquantumerror}.
The extracted thresholds are reported in Table~\ref{tab:thresholds-fit}.
For both noise models, we find thresholds in the order of $\sim10^{-2}$ and $\sim10^{-3}$, respectively, for $C=1$ and $C=d$ error correction cycles.
\\

Achieving these results requires careful management of the computational cost, which grows quickly with code distance and circuit depth.
For $C=1$, we can perform numerically exact simulations of the color code up to distance $d=7$, converging to tensor network states of moderate bond dimension ($\chi = 256$).
In contrast, the case of multiple error correction cycles $C=d$ is considerably more demanding, and for $d=7$ we were unable to reach convergence with $\chi=1024$. Consequently, the data shown in Figure~\ref{fig:Td-SRX+AD} is limited to code distances $d=3,5$.
Overall, these results show that TTN simulations can capture threshold behavior under non-Pauli noise, but the accessible regime is effectively limited to distances $d \leq 7$. In Table~\ref{tab:computetime}, we report the runtime per shot for each of the simulations that have been discussed.\\

Threshold estimation is usually performed by sampling at larger code distances, where logical error rate curves clearly intersect at a threshold point.
At smaller distances, however, finite-size effects might lead to discrepancies between crossing points obtained from different pairs of distances.
This behavior is evident in Figure~\ref{fig:T1-SRX+AD}, where the crossings between $d=3,5$ and $d=5,7$ are not consistent, and in particular, the latter is not clearly defined.
It is important to note that the fitting ansatz of Eq.~\eqref{eq:ansatz-tau} assumes an ideal scaling regime with a well-defined threshold, and therefore yields more reliable estimates when data from a broader range of code distances are available.
For this reason, the threshold values reported in Table~\ref{tab:thresholds-fit} should be interpreted, at worst, as upper bounds.
\\

\begin{table}
\def\arraystretch{1.2}
\begin{tabular}{ccc}
\hline
$\quad$ Noise model $\quad$ & $\quad C\quad$ & $\tau$ (fit) \\ \hline\hline
\multirow{2}{*}{SRX} & $1$ & $\unitsSRX=(2.56 \pm 0.01) \cdot 10^{-2}$\\
 & $d$ & $\unitsSRX=(8.62 \pm 0.69) \cdot 10^{-3}$\\\hline
\multirow{2}{*}{AD} & $1$ & $\unitsAD=(7.96 \pm 1.56) \cdot 10^{-2}$\\
 & $d$ & $\unitsAD = (5.46 \pm 0.34) \cdot 10^{-3}$\\\hline
\end{tabular}
\caption{\label{tab:thresholds-fit}\textbf{Thresholds} estimated from the ansatz of Eq.~\eqref{eq:ansatz-tau}, for each noise model, varying the number of error correction cycles $C$. Reported values include 99\% confidence intervals.}
\end{table}

\begin{table}
\def\arraystretch{1.2}
\begin{tabular}{ccccc}
\hline
Noise model & $\quad d \quad$ & $C$ & $\quad$ shot runtime (seconds) $\quad$ & $\chi$\\ \hline\hline
\multirow{2}{*}{SRX} & \multirow{2}{*}{$3$} & $1$ & 0.0495 & 16\\
 & & 3 & 0.169 & 16\\\hline
\multirow{2}{*}{SRX} & \multirow{2}{*}{$5$} & $1$ & 1.01 & 128\\
 & & 5 & 19.9 & 256\\\hline
SRX & $1$ & 1 & 56.7 & 512\\\hline\hline
\multirow{2}{*}{AD} & \multirow{2}{*}{$3$} & $1$ & 0.0578 & 8\\
 & & 3 & 0.204 & 16\\\hline
\multirow{2}{*}{AD} & \multirow{2}{*}{$5$} & $1$ & 0.317 & 16\\
 & & 5 & 3.12 & 256\\\hline
AD & 7 & 1 & 2.78 & 256\\\hline
\end{tabular}
\caption{\label{tab:computetime}\textbf{Average computational runtime} per shot for each scenario evaluated in this work. The runtimes refer to single-thread wall-time on an HPC-grade CPU. The samples have been accumulated by running code on several machines of the CINECA Leonardo (Bologna) cluster. Uncertainties on the shot runtime are to be taken at the last significant digit.}
\end{table}

\subsection{\label{ssec:twirl}Pauli twirling comparison}

Finally, we compare the results of our simulations with those obtained under the Pauli twirling approximation for the noise models considered in this work~\cite{PaulinTNsurface}.
Let $\mathcal{E}[\rho] = \sum_i K_i \rho K_i^\dag$ be an arbitrary quantum channel, which we express in the Pauli operator basis as
\begin{equation*}
    \mathcal{E}[\rho]=\sum_{i,j} \eta_{i,j} P_i \rho P_j \quad,
\end{equation*}
where $P_i$ and $P_j$ are Pauli operators. Notice that expressing the Kraus operators in the Pauli basis will introduce, in general, off-diagonal terms $\eta_{i,j}$, $i\neq j$.
The Pauli twirling approximation~\cite{PhysRevA.87.030302,PhysRevA.87.012324,katabarwa2017dynamicalinterpretationpaulitwirling} is obtained by removing the off-diagonal terms of $\eta$, yielding a Pauli channel that can be efficiently simulated by sampling in Clifford simulators:
\begin{equation}
    \mathcal{E}^{\mathrm{(PTA)}}[\rho] = \sum_i \eta_{i,i} P_i \rho P_i \quad.
\end{equation}

In Ref.~\cite{PaulinTNsurface}, the AD model showed good agreement with the Pauli twirling approximation, while a coherent noise model (specifically, a coherent Z-axis rotation) deviated significantly from its twirled counterpart.

\begin{figure}
    \centering
    \includegraphics[width=\linewidth]{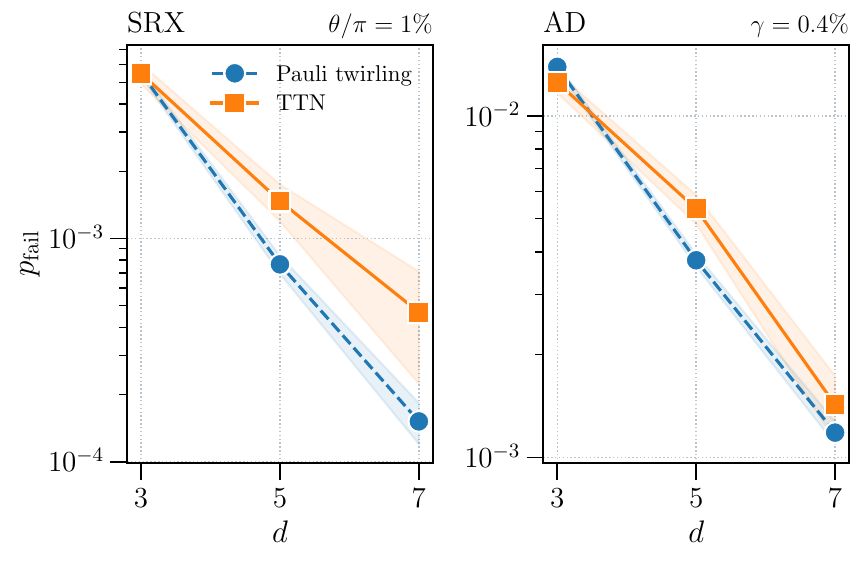}
    \caption{\textbf{Comparison with Pauli Twirling Approximation of the noise models.} 
    We simulate a single round of error correction ($C=1$) at fixed noise strengths: the SRX model (left) with $\unitsSRX = 10^{-2}$ and the AD model (right) with $\unitsAD=4\cdot10^{-3}$. 
    In both cases, we compare the logical failure rate obtained from TTN simulations with that predicted by the Pauli twirling approximation, evaluated using a Clifford simulator. 
    The shaded area represents 99\% confidence interval.
    The results show good agreement between the exact and twirled models for AD, while for SRX the discrepancy grows with code distance.}
    \label{fig:twirl-comparison}
\end{figure}

We compare the logical error rate at increasing distance of the code for a fixed noise strength. Our results are consistent with previous findings.
In line with Ref.~\cite{PaulinTNsurface}, we observe that the AD model matches well with its Pauli twirling approximation. 
In contrast, the SRX model - based on coherent X-axis rotations - shows only partial agreement with the Pauli-twirled approximation. While the two methods yield similar results for $d=3$, their predictions diverge as the code distance increases.
This behavior, illustrated in Figure~\ref{fig:twirl-comparison}, points out that the accuracy of the twirling approximation is not straightforward to predict. Direct simulations with alternative methods, such as our TTN approach, are therefore essential to assess its reliability.
Furthermore, whereas Ref.~\cite{PaulinTNsurface} reported discrepancies only when noise was applied to an already prepared logical state, our simulations demonstrate that such deviations persist under circuit-level noise and multiple rounds of syndrome extraction. 
This numerical evidence supports previous studies, such as Ref.~\cite{PhysRevA.91.022335}, where the deviation has been observed for the Steane code (the $d=3$ color code).

\section{\label{sec:conclusion}Conclusions}

In this work, we have presented a tensor network simulation framework for analyzing quantum error correction protocols under realistic, circuit-level noise models.
Specifically, we have performed simulations on a Tree Tensor Network (TTN) ansatz to estimate code thresholds under two different noise models: a coherent single-qubit rotation noise (SRX) and an amplitude damping (AD) model.
Our results demonstrate that the TTN-based approach can simulate full circuit-level noise models for quantum error correction codes at nontrivial distances, with sufficient accuracy to extract meaningful thresholds.
We have also shown that Pauli approximations of the noise models lead to incorrect estimation of the logical failure rate, motivating the need for alternative simulation strategies.

These simulations highlight both the potential and the limitations of TTN-based quantum error correction simulations. In particular, careful control of the bond dimension and explicit validation against approximation artifacts are essential to ensure reliable results.\\

Looking forward, tensor networks could be applied to a broader range of problems beyond threshold estimation with non-Pauli noise models.
First, tensor networks provide native access to quantities of central importance in quantum information theory, such as entanglement~\cite{eisert2013entanglementtensornetworkstates} and magic monotones~\cite{Haug2023stabilizerentropies}, paving the way for threshold estimation based on coherent information~\cite{Colmenarez_2024}.
Second, since non-Clifford operations do not pose a significant obstacle for tensor network simulations, this approach is well-suited for simulating non-Clifford protocols in quantum error correction, such as magic state injection~\cite{rodriguez2024experimentaldemonstrationlogicalmagic}, which remain inaccessible to traditional stabilizer simulators.
Finally, future work will depend on optimizing tensor network ansätze to represent stabilizer states more efficiently.
This research direction has been recently pursued, either by constructing efficient tensor network representations of stabilizer states~\cite{PhysRevLett.133.230601,Nakhl_2025} or by introducing Clifford disentanglers~\cite{Mello_2025,reinić2025augmentedtreetensornetwork}. Such techniques mitigate the rapidly increasing computational cost associated with simulating higher-distance codes or deeper error correction circuits, where entanglement growth becomes a dominant bottleneck.

\begin{acknowledgments}
We acknowledge Markus Müller for feedback on this work.
The research leading to these results has received funding from the following organizations: Italian Research Center on HPC, Big Data and Quantum Computing (NextGenerationEU Project No. CN00000013), project EuRyQa (Horizon 2020), project PASQuanS2 (Quantum Technology Flagship); Italian Ministry of University and Research (MUR) via: Quantum Frontiers (the Departments of Excellence 2023-2027); German Federal Ministry of Education and Research (BMBF): project QRydDemo (the funding program quantum technologies - from basic research to market); the World Class Research Infrastructure - Quantum Computing and Simulation Center (QCSC) of Padova University; Istituto Nazionale di Fisica Nucleare (INFN): iniziativa specifica IS-QUANTUM. We acknowledge computational resources from Cineca on the Leonardo machine and the INFN Padova HPC cluster.
\end{acknowledgments}

\appendix

\section{\label{apx:qec}Generic elements of Quantum Error Correction}

For completeness, we provide here a more detailed overview of the quantum error correction concepts briefly introduced in Section~\ref{sec:problem}.

\subsection{\label{apx:qec-colorcode}The topological formulation of the color code}

\begin{figure}
    \centering
    \includegraphics[width=\linewidth]{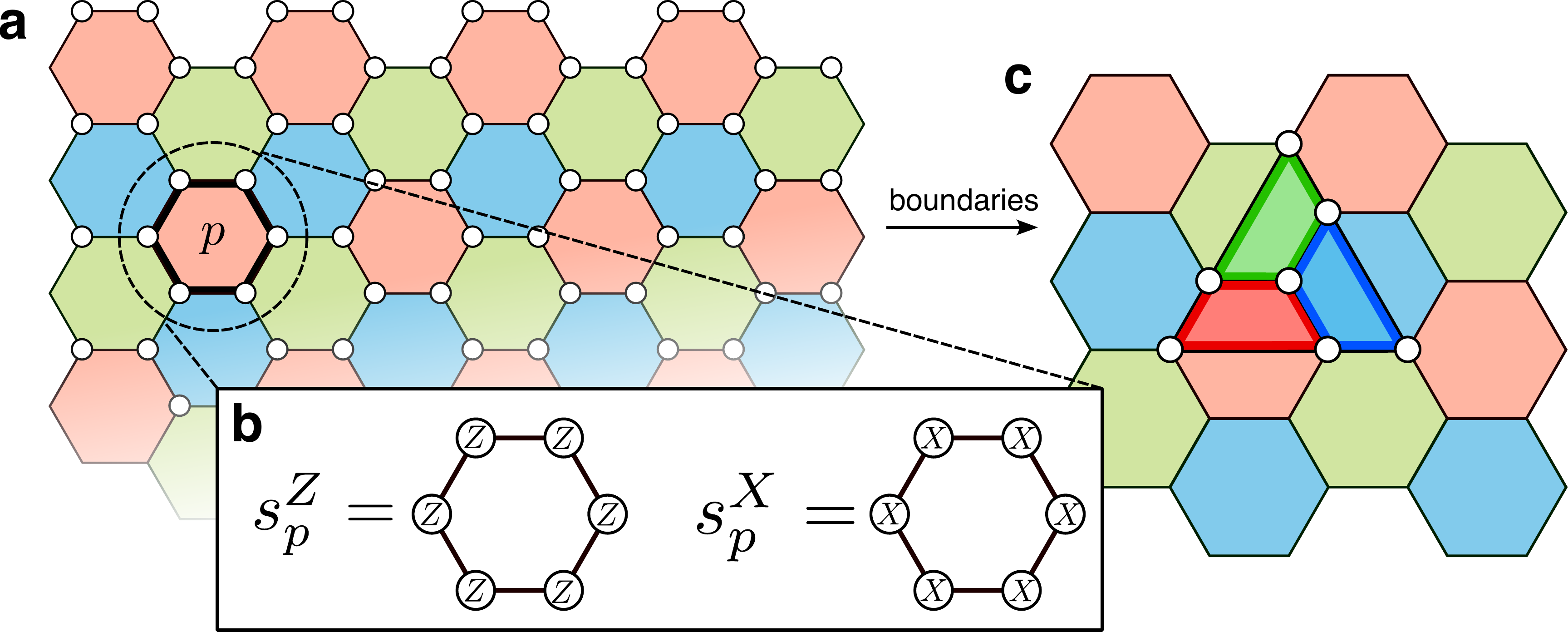}
    \caption{\textbf{The color code.} \textbf{(a)} Honeycomb lattice with qubits placed on its vertices. A plaquette $p$ is highlighted. \textbf{(b)} The stabilizer operators of the color code acting on $p$. \textbf{(c)} Color code on a finite patch of the hexagon lattice. The cut shown in the figure is a $[[7,1,3]]$ code, the smallest in the color code family. Notice that the stabilizers on the boundary qubits act on 4 qubits rather than 6.}
    \label{fig:colorcode}
\end{figure}

In the honeycomb color code, stabilizers are defined on a two-dimensional lattice of hexagons, where physical qubits reside on the vertices of the lattice~\cite{PhysRevLett.97.180501,Kubica_2015} (see Figure~\ref{fig:colorcode}a). For each face of the hexagonal lattice, called a plaquette $p$, two stabilizers are associated: $s_p^X$, a tensor product of Pauli-$X$ operators, and $s_p^Z$, a tensor product of Pauli-$Z$ operators, both acting on the qubits at the vertices of $p$ (see Figure~\ref{fig:colorcode}b).

The honeycomb lattice is three-colorable, meaning that one can assign three distinct colors (typically red, green, and blue) to the plaquettes such that two adjacent faces never share the same color. This coloring underpins many features of the code, including its logical operators and decoding strategies.\\

The color code recipe is highly flexible. When qubits are placed on a topologically nontrivial manifold, the color code can be employed within a topological error correction framework~\cite{bombin2013introductiontopologicalquantumcodes}. 
In this context, stabilizers correspond to six-body interactions, and the code space is identified with the degenerate ground-state manifold of a gapped Hamiltonian. Logical qubits are then protected by topological order, with a finite energy gap suppressing local excitations (errors).
\\

Alternatively, one can restrict the 2D lattice to a finite triangular patch, producing a planar code with boundaries.
The minimal such region is illustrated in Figure~\ref{fig:colorcode}c.
Then, the key idea is to arrange the $n$ physical qubits into a quantum circuit and perform repeated measurements of the stabilizer operators. 
These stabilizer measurements are non-destructive, as they preserve the encoded logical information while yielding information about potential physical errors.

The outcome of each stabilizer measurement is a classical bit, known as the \emph{syndrome}. 
A $+1$ outcome indicates that the system remains in the +1 eigenspace of the corresponding stabilizer, while any $-1$ outcome signals the presence of an error that has taken the system out of the code space.

We now focus on a finite patch of the honeycomb lattice, such as the triangular layout illustrated in Figure~\ref{fig:colorcode}c. Due to the presence of boundaries, the color code encodes a single logical qubit, i.e., $k=1$. 
The stabilizer operators retain their structure as Pauli-$X$ and Pauli-$Z$ strings over each plaquette; however, their weight is reduced near the boundaries, from six in the bulk to four along the edges.
A standard method for measuring the syndromes assigns one ancilla qubit per stabilizer and performs a sequence of CNOT gates between the ancilla and the data qubits on the corresponding plaquette. Figure~\ref{fig:circuitlevel}a shows the quantum circuits required to measure both $s_p^X$ and $s_p^Z$ stabilizers.
This procedure results in an overhead of two ancilla qubits per plaquette (one for each stabilizer type).
To render syndrome extraction fault-tolerant and prevent error propagation, alternative protocols require an additional ancilla overhead due to flagged measurements~\cite{Chamberland2018flagfaulttolerant,Chamberland_2020b,Chamberland_2020}.
Another line of research explores the removal of measurements altogether via measurement-free error-correction schemes~\cite{PRXQuantum.5.010333}.
Such schemes, however, lie beyond the scope of this work, which is to assess the performance of tensor network simulations for circuit-level noise models in a minimal and controlled setting.\\

\subsection{\label{apx:qec-decoding}Decoding}

Decoding strategies are very often adapted from classical error correction algorithms, but they must be modified to account for the algebraic structure specific to quantum codes.
Decoding strategies for QECCs are typically based on approximate inference of the error, with the most common approaches being Most Likely Error (MLE) and Maximum Likelihood Decoding (MLD)~\cite{Sundaresan_2023,lacroix2024scalinglogiccolorcode}.
While MLD is optimal - hence, it leads to optimal threshold estimations - it relies on solving a \#P-complete problem~\cite{7097029,9456887}, which motivates the search for more efficient decoding schemes.\\

In this work, we use another decoding algorithm~\cite{Lee2025colorcodedecoder}, specifically studied for color code.
Minimum-Weight Perfect Matching (MWPM)~\cite{Higgott_2025} interprets non-trivial syndrome measurements as pairs of defects on a lattice. Then, the decoding is reformulated as a graph matching problem: identifying a minimum-weight error configuration that links the defects along the shortest path under the constraints of the lattice geometry.
MWPM is particularly well-suited to surface codes~\cite{bravyi1998quantumcodeslatticeboundary}, where Pauli-string errors produce naturally pairs of defects~\cite{chandra2025distributedrealizationcolorcodes}.
However, applying MWPM to color codes is non-trivial, because there exist syndrome patterns that cannot be matched pairwise. For instance, a bitflip error on a single qubit anticommutes with 3 stabilizer checks. As a result, naive MWPM decoding strategies for color codes show suboptimal scaling of the logical failure rate~\cite{PRXQuantum.2.020341,PhysRevResearch.6.033086}.\\

The \emph{concatenated MWPM decoder} introduced in Ref.~\cite{Lee2025colorcodedecoder} addresses these limitations through a two-step procedure:
\begin{enumerate}
    \item MWPM can be applied to the color code by considering defects on a restricted set of symmetries~\cite{PhysRevA.89.012317}. 
    
    Let us label each stabilizer by the color of the plaquette $p$ on which it acts. Let $\sigma_{c} = \{s_p^i : \mathrm{color}(p) = c,\; i=X,Z \}$, where $c \in \{\text{red}, \text{green}, \text{blue}\}$.
    Now define a reduced stabilizer set by omitting one specific color, say $\Sigma_{\mathrm{red}}=\sigma_{\mathrm{blue}}\cup\sigma_{\mathrm{green}}$.
    Then, the key observation is that the stabilizers in $\Sigma_{\mathrm{red}}$ will produce an even number of defects on the red faces~\cite{PRXQuantum.3.010310}. The same consideration applies to $\Sigma_{\mathrm{blue}}$ and $\Sigma_{\mathrm{green}}$. 
    This suggests that MWPM can be applied to color codes by restricting to a sublattice of all colors but one.
    \item After decoding on a reduced lattice $\Sigma_{c}$, a set of predicted error edges is obtained for color $c$. These edges, together with the syndromes of the stabilizers $\sigma_{c}$, define a new decoding problem, which is also solved via MWPM. The outcome of this second step is a predicted error configuration.
\end{enumerate}
The procedure above is repeated for all three colors, yielding three candidate corrections. The most likely guess (the lowest weight error) is finally chosen as the algorithm output.
For a full discussion of the decoder design and implementation, we refer to the original publication~\cite{Lee2025colorcodedecoder}.

\subsection{\label{apx:qec-threshold}Threshold estimation}

A key result in quantum error correction theory is the \emph{threshold theorem}~\cite{knill1996thresholdaccuracyquantumcomputation}. 
Intuitively, the theorem states that if the noise strength $p_{\mathrm{err}}$ is below a critical value - the \emph{threshold} $\tau$ - then the QECC can correct logical errors at a rate faster than they accumulate.
Here, $p_{\mathrm{err}}$ represents the noise strength under a given error model, which has been discussed in §\ref{ssec:circlevel}.
Determining the threshold $\tau$ is essential, as it drives both the theoretical design and experimental validation of quantum hardware.
It also serves as a benchmark for comparing the performance of different QECCs.
However, the threshold is highly sensitive to the details of the system, depending on the modeling of the noise that is acting on the qubits, the algorithm used to decode the syndrome, and even on how the syndrome extraction and logical operations are executed.\\

In the most general settings, thresholds are estimated numerically via a Monte Carlo sampling~\cite{PhysRevA.89.022321,landahl2011faulttolerantquantumcomputingcolor}. These simulations sample the logical failure rate $p_{\mathrm{fail}}$ by repeatedly executing a representative quantum error correction circuit under noise.
The key ingredients of this procedure are the following: (i) choose a circuit of the QECC and fix its parameters; (ii) sample errors from a model and insert them in the circuit; (iii) run the simulation and decode the syndrome.
Steps (ii) and (iii) are repeated multiple times to estimate $p_{\mathrm{fail}}$, i.e., the ratio between decode failures and the total number of simulations, with acceptable statistical uncertainty.
To extract a threshold from these simulations, steps (i) and (ii) are systematically varied to explore different code distances and noise strengths. 
The logical error rate typically follows a scaling relation of the form $p_{\mathrm{fail}} \sim (p_{\mathrm{err}}-\tau)^{\alpha d + \beta}$~\cite{Watson_2014}. The threshold can then be estimated by fitting this scaling behavior to simulation data obtained from different values of $d$.\\

Assuming the complexity of the decoder is negligible, the major overhead in this process is the circuit simulation of step (iii).
As it is standard in the quantum error correction field, these simulations are typically performed using Clifford simulators.
Although this may seem like a strong limitation, the use of Pauli noise models remains theoretically well-motivated.
In stabilizer codes, all errors are effectively projected onto Pauli strings during syndrome extraction, so the ability to correct arbitrary Pauli-group errors up to some weight guarantees correctability of more general quantum error~\cite{gottesman2009introductionquantumerrorcorrection}.

That said, the noise models themselves can model non-Pauli errors, but the syndrome extraction process forces every error to collapse into a string of Pauli operators on the code qubits~\cite{landahl2011faulttolerantquantumcomputingcolor}.
The catch is that linking general error processes to the syndrome patterns they induce (in terms of Pauli-group errors) is highly non-trivial. Consequently, a valid assessment of the threshold under generic noise may require simulation methods not restricted to the Clifford group.

\section{\label{apx:layout}Optimal qubit ordering on binary TTN}

\begin{figure}
    \centering
    \includegraphics[width=\linewidth]{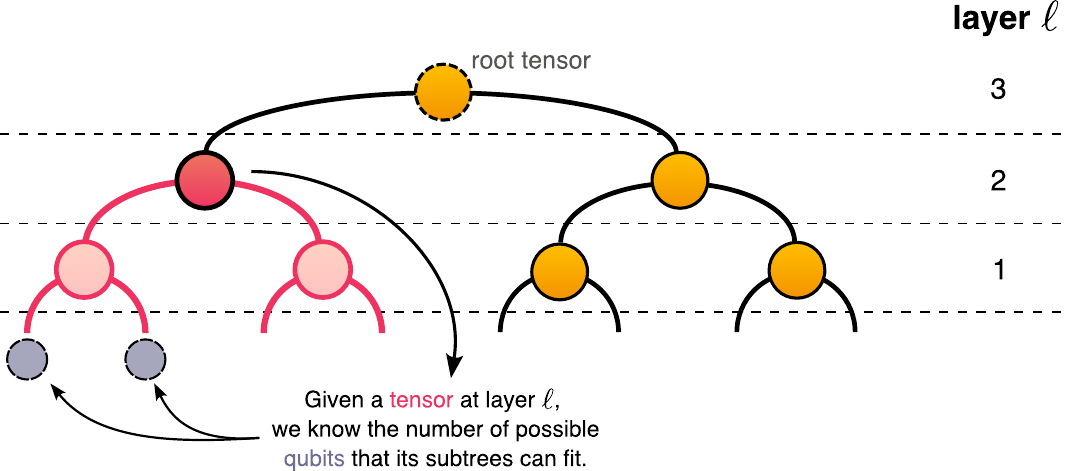}
    \caption{\textbf{Schematic structure of a binary TTN.} The diagram follows standard tensor network notation: circles denote tensors, and edges represent their indices. Edges that are not shared between two tensors correspond to free indices and are arranged along the bottom layer of the network. These free indices represent the physical degrees of freedom of the system - in this case, the qubits.
    In our implementation, the root tensor is omitted in the simulation, as it can be trivially absorbed into one of the two tensors directly below it.
    The tensors are organized in layers labeled by $\ell$, with the numeration proceeding in descending order from the root of the tree. This convention allows one to determine the number of leaves, i.e., qubits, that each subtree at level $\ell-1$ can accommodate: $2^{\ell-1}$.
    The depth of a TTN with $\ell$ layers is $\ell$, by definition.
    }
    \label{fig:ttn-layout}
\end{figure}

\begin{figure*}
    \centering
    \includegraphics[width=\linewidth]{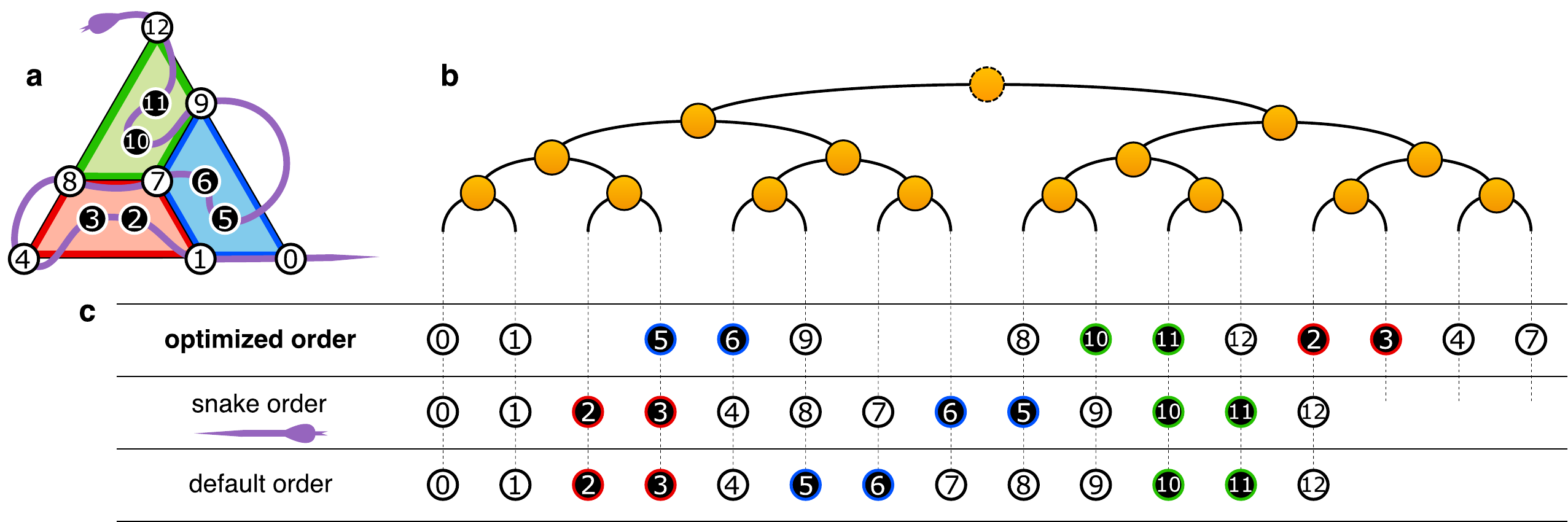}
    \caption{\label{fig:d3clustering}\textbf{Example of optimized qubit ordering on TTN for the distance-3 color code.} 
    \textbf{(a)} The $[[7,1,3]]$ color code. White circles denote the data qubits, while black circles represent ancilla qubits used to measure stabilizers associated with each plaquette. Qubits are labeled from 0 to 12. We define the \textit{default} ordering as the one that follows this labeling, leaving the empty sites of the TTN to the right. The \textit{snake} ordering is illustrated by the purple path connecting the qubits.
    \textbf{(b)} A complete binary TTN with 16 physical sites, which is the smallest such TTN capable of accommodating the $N=13$ qubits of the $[[7,1,3]]$ code. Physical sites must be assigned to qubits, although some sites may remain unused. The yellow circles represent the tensors, each having an upper bond, connected to a tensor of the upper layer, and two downward bonds.
    \textbf{(c)} Qubit-to-site assignments for three layouts: the \textit{optimized} order computed by our algorithm, the \textit{snake} order of panel (a), and the \textit{default} order based on qubit labels.
    }
\end{figure*}

We now address the problem of assigning the qubits of a quantum circuit to the leaves of a complete binary TTN.
A central challenge in tensor network-based simulation of quantum circuits lies in faithfully capturing the entanglement structure of the quantum state being represented.
A poor mapping of qubits to the physical indices of a tensor network ansatz may place highly entangled qubits far apart in the network, requiring a significantly larger bond dimension to preserve the relevant correlations, thereby increasing simulation costs.
This problem is relevant for the MPS, where the tensors are placed in a linear layout, as well as for the TTN ansatz, where the tensors are arranged in a tree network, like the one in Figure~\ref{fig:ttn-layout}.

Ref.~\cite{Seitz_2023} introduced a heuristic method to construct a tree topology based on the structure of quantum gates acting on pairs of qubits. 
However, that approach is agnostic to user-defined constraints on the tree topology. 
In our work, we are restricted by the simulation library - Quantum Matcha Tea~\cite{qmatchatea,qtealeaves,qredtea} - to use a complete binary TTN layout.
This constraint, however, does not necessarily imply a loss of efficiency: an entanglement pattern captured by a general tree can, in principle, be replicated in a binary tree by suitably increasing its depth and carefully reassigning qubits to its leaves.
\\

Inspired by Ref.~\cite{Seitz_2023}, we adopt a clustering-based strategy and formulate a new algorithm (Algorithm~\ref{alg:tree_structure_binary}) to construct the binary TTN layout from top to bottom.
Starting from the root tensor of the tree, the algorithm recursively splits the qubit set into two subgroups, assigning each to one of the two child subtrees. 
Knowing the current level of the tree $\ell$, we can determine the maximum number of qubits that each subtree can accommodate: $2^{\ell-1}$. 
To guide the split, we employ a clustering procedure that attempts to partition the qubits into two clusters based on their mutual connectivity.
If the resulting clusters do not fit within the subtree capacity constraints, the algorithm iteratively increases the number of clusters, then merges them until a valid binary split is found.

The clustering step can, in principle, use any standard clustering algorithm, but we follow Ref.~\cite{Seitz_2023} and employ spectral clustering with a custom similarity function.
Let us call $q_i$ the $i^{th}$ qubit of the input circuit, where $i=1,\dots,N$.
Then, define $G(q_i)$ as the set of multi-qubit gates acting on qubit $q_i$.
The custom similarity function is defined as
\begin{equation}
    f_{qc}(q_i,q_j) = |G(q_i)\cap G(q_j)| + \frac{1}{|G(q_i)|+|G(q_j)|}
\end{equation}
for $i\neq j$. This function promotes the grouping of qubits that frequently interact, while penalizing isolated qubits.
\\

Given the heuristic nature of the clustering-based layout construction, convergence to an optimal qubit-to-site assignment is not guaranteed. 
Consequently, it may be necessary to generate multiple candidate layouts and select the most promising one. 
This selection can be performed by either executing short simulations and evaluating the resulting maximum bond dimension $\bdexact$, or through alternative cost heuristics that estimate layout quality without requiring full simulation, though we do not explore such heuristics in this work. 
For our purposes, we found it sufficient to sample a small number of candidate layouts from the clustering algorithm and post-select the one that minimized $\bdexact$.
\\

\begin{algorithm}
    \newcommand\mycommfont[1]{\footnotesize\ttfamily{#1}}
    \SetCommentSty{mycommfont}
    \DontPrintSemicolon
    \SetKwProg{Fn}{Function}{:}{}
    \SetKw{KwTo}{in}
    \SetKwFunction{c}{cluster}\SetKwFunction{func}{find\_tree\_structure}\SetKwFunction{cs}{create\_subtree}\SetKwFunction{Node}{Node}\SetKwFunction{csm}{similarity\_matrix}\SetKwData{sim}{similarity}\SetKwData{nc}{c}\SetKwData{clab}{cluster\_labels}\SetKwData{qc}{qc}\SetKwData{dmax}{$D$}\SetKwData{q}{qubits}
    \Fn{\func{\qc, \dmax}}{
        \KwData{\qc = Quantum\ Circuit, \dmax = Target\ tree\ depth}
        \KwResult{Tree structure}
        \tcc{Check \dmax is sufficient to accommodate all the qubits in \qc} 
        \q $\leftarrow$ \qc.qubits\;
        \Return{\cs{\q, \dmax}}
    }
    \BlankLine\BlankLine
    \SetKwData{ll}{label}\SetKwData{qsub}{qubit\_subset}\SetKwData{children}{children}\SetKwData{sim}{sim}\SetKwData{cl}{layer}
    \SetKwFunction{ca}{clustering\_algorithm}
    \SetKwFunction{fit}{fits\_in\_subtrees}
    \SetKwFunction{res}{random\_even\_split}
    \SetKwFunction{red}{reduce\_to\_two\_clusters}
    \Fn{\cs{\q, \cl}}{
        \KwData{\q= list\ of\ unassigned\ qubits, \cl= current\ layer}
        \KwResult{Tree structure}
        \BlankLine
        \tcc{Handle trivial cases first}
        \uIf{\cl = 1}{
            \children $\leftarrow$ $\{ q$ \textbf{foreach} $q$ \textbf{in} $\q\}$\;
            \Return{\Node{\children}}
        }

        \uIf{$|\q| \le$ 2}{
            \children $\leftarrow$ $\{$ \cs($q$, \cl-1) \textbf{foreach} $q$ \textbf{in} \q $\}$\;
            \Return{\Node{\children}}
        }
        \BlankLine
        \tcc{Start clustering}
        \For{$k\in[2,|\q|]$}{
            \clab $\leftarrow$ \c{\q, $k$}\;
            \clab $\leftarrow$ \red{\clab}\;
            \uIf{\fit{\clab, \cl-1}}{
                \textbf{break}\;
            }
        }
        \BlankLine
        \tcc{In case all clustering attempts have failed, splits the qubits in two random clusters.}
        \uIf{$\max(\clab) \ge$ 3}{
            \clab $\leftarrow$ \res{\q}
        }
        \BlankLine
        \tcc{Create subtrees according to the labels}
        \children $\leftarrow \emptyset$\;\label{marker}
        \ForEach{\ll \KwTo \clab }{
            \qsub $\leftarrow$ $\q[\ll]$ \;
            \children $\leftarrow$ \children $\cup$ \cs{\qsub, $ \cl-1$}\;
        }
        \Return{\Node{\children}}
    }
    \BlankLine\BlankLine
    \Fn{\c{\q, k}}{
        \KwData{\q= list\ of\ unassigned\ qubits, k= number\ of\ clusters}
        \KwResult{list of cluster labels $(1,2,\dots)$ }
        \sim $\leftarrow$ \csm{\q}\tcp*{$f_{\text{qc}}$}
        \Return{\ca{\sim, k}}\tcp*{any clustering, e.g., SpectralClustering}
    }
    \BlankLine
    \Fn{\fit{\clab, \cl}}{
        \KwResult{True or False}
        \tcc{Check if the clusters defined by \clab fit into two subtrees, which can have at most $2^{\cl}$ leaves.}
    }
    \BlankLine
    \Fn{\red{\clab}}{
        \KwResult{list of two cluster labels $(1,2)$}
        \tcc{This function agglomerates cluster labels, aiming to create two balanced clusters.}
    }
    \caption{Binary tree structure search}
    \label{alg:tree_structure_binary}
\end{algorithm}

The benefit of optimizing the qubit ordering becomes evident when examining the data presented in Table~\ref{tab:bdexact}. 
For small circuits, such as the one corresponding to code distance $d=3$, the improvement introduced by the optimized layout is marginal. 
However, as the number of qubits $N$ increases with higher-distance codes, the difference between the optimized and the snake ordering becomes substantial, leading to significant savings in computational resources.
The advantage is further amplified when simulating deeper circuits - e.g., for $C=d$ error correction cycles - where the required bond dimension $\bdexact$ grows and eventually saturates to the values shown in the table. 
In these regimes, the optimized layout consistently yields more efficient tensor network representations. 
Furthermore, this optimization becomes essential for simulating circuits at distance $d=7$, where exact simulations, i.e., without bond dimension truncation, would be computationally prohibitive in absence of an optimized qubit mapping.
An example of optimal mapping for the distance-3 color code is reported in Figure~\ref{fig:d3clustering}.

\begin{table}
\def\arraystretch{1.3}
\begin{tabular}{cccccc}\hline
 & & & \multicolumn{3}{c}{$\chi$ for qubit ordering} \\ \cline{4-6} 
$\quad d\quad$ & $\quad N\quad$ & $\quad C\quad$ & default & snake & \textbf{optimized}\\ \hline\hline
\multirow{2}{*}{3} & \multirow{2}{*}{13} & 1 & 8 & 8 & 8 \\
 & & 3 & 32 & 16 & 16 \\\hline
\multirow{2}{*}{5} & \multirow{2}{*}{37} & 1 & 512 & 256 & 16 \\
 & & 5 & 2048 & 1024 & 256 \\\hline
\multirow{2}{*}{7} & \multirow{2}{*}{73} & 1 & $\ge$2048 & 2048 & 256 \\
 & & 7 & - & - & $\ge$2048 \\ \hline
\end{tabular}
\caption{\label{tab:bdexact}\textbf{Bond dimensions $\bdexact$ for different TTN qubit orderings in simulations of the AD noise model.} 
The \textit{default} ordering assigns qubit $i$ to TTN site $i$. Figure~\ref{fig:d3clustering} illustrates examples of the \textit{snake} and \textit{optimized} orderings for the $d=3$ color code. 
We observe that the optimized ordering provides a significant advantage, especially when simulating color codes at higher distances. 
The required bond dimension $\bdexact$ also increases with the circuit depth, i.e., with the number of error correction cycles $C$.
To contextualize the computational effort, we report the total number of qubits $N$ for each simulated code distance.
}
\end{table}

\clearpage
\bibliography{apssamp}

\end{document}